\documentclass[11pt,a4paper]{article}
\usepackage[hyperref]{naaclhlt2018}
\usepackage{times}
\usepackage{latexsym}

\usepackage{url}
\usepackage{graphicx}

\usepackage[utf8]{inputenc} 
\usepackage[T1]{fontenc}    
\usepackage{hyperref}       
\usepackage{url}            
\usepackage{relsize}        
\usepackage{booktabs}       
\usepackage{amsfonts}       
\usepackage{nicefrac}       
\usepackage{microtype}      
\usepackage{comment}        
\usepackage{amsopn}
\usepackage{xcolor}
\usepackage{multirow}

\aclfinalcopy 

\usepackage{todonotes}
\newcounter{hfang}

\newcounter{chenghao}

\newcounter{eclark}

\newcounter{maarten}

\newcounter{ari}

\newcounter{mari}

\newcounter{noah}

\title{Sounding Board: A User-Centric and Content-Driven Social Chatbot}

\author{Hao Fang$^*$ \quad {\bf Hao Cheng}$^*$ \quad
	{\bf Maarten Sap}$^\dag$
	\quad {\bf Elizabeth Clark}$^\dag$ \quad {\bf Ari Holtzman}$^\dag$ \\
	\quad {\bf Yejin Choi}$^\dag$ \quad 
	{\bf Noah A. Smith}$^\dag$ 
		\quad {\bf Mari Ostendorf}$^*$ \\
		University of Washington \\
	$^*$\texttt{\{hfang,chenghao,ostendor\}@uw.edu} \\
	$^\dag$\texttt{\{msap,eaclark7,ahai,yejin,nasmith\}@cs.washington.edu}
}

\begin{document}
\maketitle
\begin{abstract}
We present 
Sounding Board, 
a social chatbot that won the 2017 Amazon Alexa Prize.
The system architecture 
consists of several components including 
spoken language processing, 
dialogue management, language generation, and content management,  
with emphasis on \emph{user-centric} and \emph{content-driven} design. We also share insights gained from large-scale online logs based on 160,000 conversations with real-world users. 
\end{abstract}

\section{Introduction}
\label{sec:intro}
Researchers in artificial intelligence (AI) have long been interested in the challenge of developing a system that can have a coherent conversation with humans: early systems include Parry \cite{Colby1975AI}, Eliza \cite{Weizenbaum1996Comm}, and Alice \cite{Wallace2009ALICE}.
The Loebner Prize\footnote{\url{http://aisb.org.uk/events/loebner-prize}} is a Turing test for conversational AI that has challenged researchers since 1990.
Recent work has addressed tasks where passing the Turing test is not a concern.
Goal-oriented conversational systems facilitate natural user interaction with devices via text and spoken language. These AI assistants typically focus on short interactions, as in commercial products such as Amazon Alexa, Microsoft
Cortana,  Google Assistant, and Apple Siri. General conversational systems, called chatbots, have constrained social interaction capabilities but have difficulty
generating conversations with long-term coherence
\citep{serban2017hierarchical,sato2017modeling,shao2017generating,tian2017make,Ghazvininejad2018AAAI}.

The Alexa Prize sets forth a new challenge: creating a system that can hold a
coherent and engaging conversation on current events and popular topics such as sports,
politics, entertainment, fashion and technology
\cite{AlexaPrize2017}.
Our system, {Sounding Board},\footnote{\url{https://sounding-board.github.io}}
demonstrates that it is feasible to build an agent that can engage in long-term conversation when backed by rich content and knowledge of the user obtained through interaction.
Sounding Board won the inaugural Amazon Alexa 
Prize with an average score of 3.17 on a 5-point scale  and an average conversation duration of 10:22, evaluated by a panel of independent judges.\footnote{\url{https://developer.amazon.com/alexaprize/2017-alexa-prize}}

\begin{figure}[t]
	\centering 
	\vspace{-4mm}
	\includegraphics[width=0.46\textwidth,trim=2cm 9cm 3.5cm 1cm]{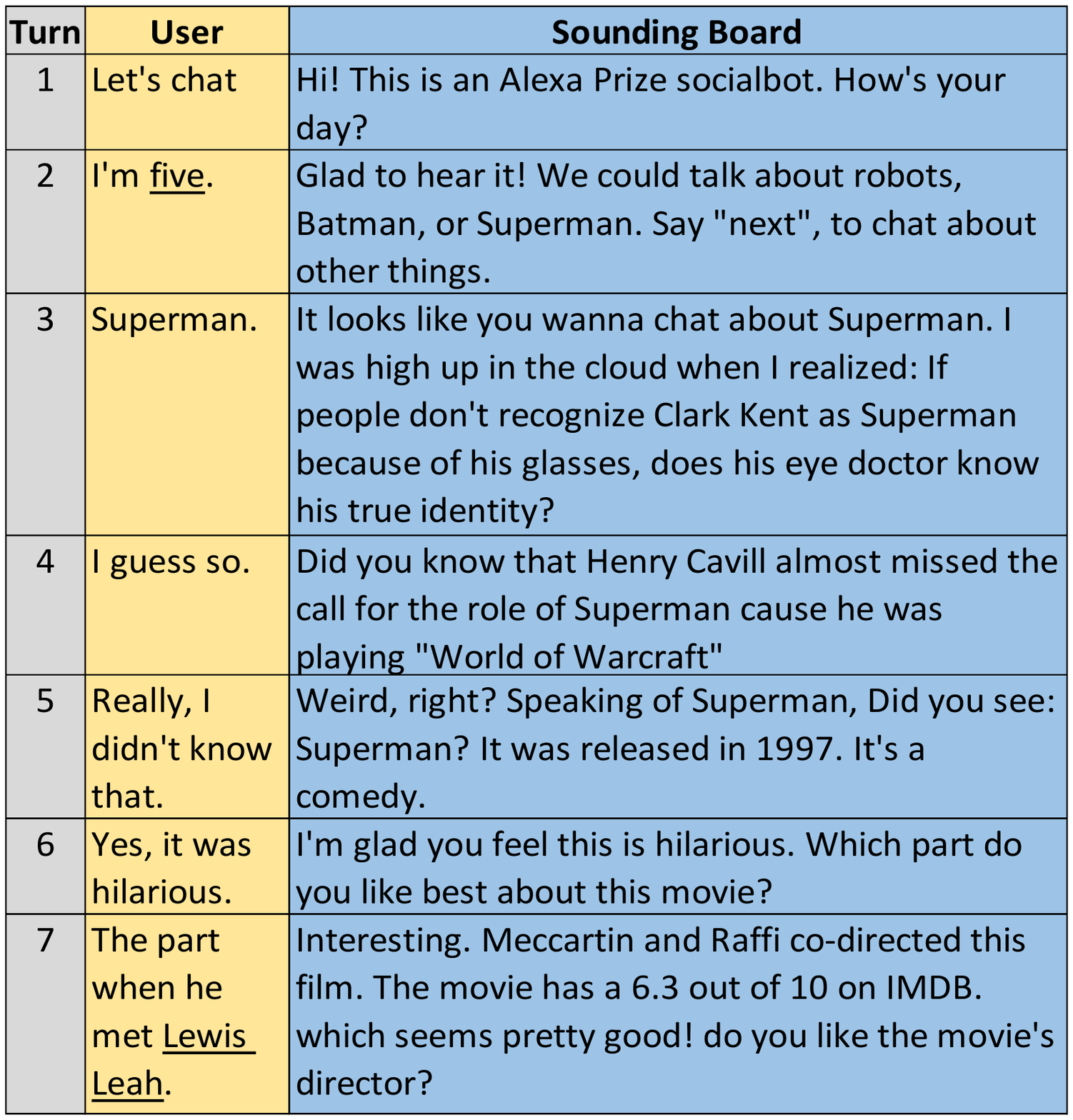}
	\caption{A sample dialog. 
	Suspected speech recognition errors in the user utterances are underlined.}
	\label{fig:sample_convo}
\end{figure}

There are two key design objectives of Sounding Board: to be \emph{user-centric} and \emph{content-driven}. 
Our system is \textit{user-centric} in that 
users can control the topic of conversation, while the system adapts responses to the user's likely interests by gauging the user's personality.
Sounding Board is also \emph{content-driven}, as it continually supplies interesting and relevant information to continue the conversation, enabled by a rich content collection that it updates daily.
It is this content that can engage users for a long period of time and drive the conversation forward. A sample conversation is shown in Fig.~\ref{fig:sample_convo}. 

We describe the system architecture in \S\ref{sec:system_arch}, share our insights based on  large scale conversation logs in \S\ref{sec:insights}, and conclude in \S\ref{sec:conclusion}.

\section{System Architecture}
\label{sec:system_arch}
\begin{figure*}[t]
	\centering 
	\includegraphics[width=0.8\textwidth,trim=1cm 1.2cm 1cm 0.8cm]{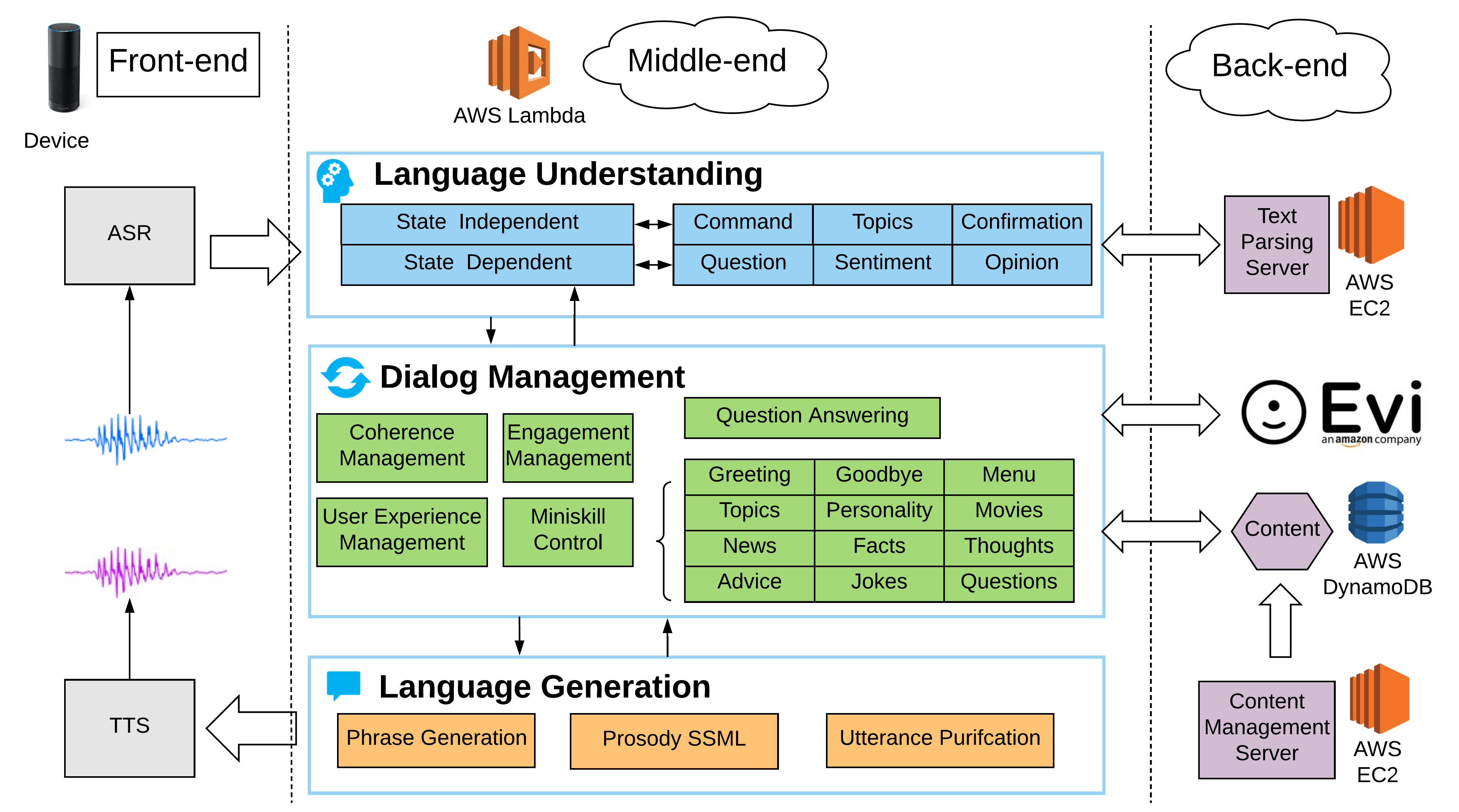}
	\caption{System architecture.
		{\bf Front-end}: Amazon's Automatic Speech Recognition (ASR) and Text-to-Speech
	(TTS) APIs.
	{\bf Middle-end}: NLU, DM and NLG modules implemented using the AWS Lambda service. 
	{\bf Back-end}: External services and AWS
	DynamoDB tables for storing the knowledge graph.}
	\label{fig:system_arch}
\end{figure*}

Sounding Board uses a modular framework as shown in
Fig.\,\ref{fig:system_arch}.
When a user speaks, the system produces a response using three modules:
natural language understanding (NLU),
dialog manager (DM),
and natural language generation (NLG).
The NLU produces a representation of the current event by analyzing the user's speech given the current dialog state (\S\ref{ssec:nlu}).
Then, based on this representation, the DM executes the dialog policy and
decides the next dialog state (\S\ref{ssec:dm}). 
Finally, the NLG uses the content selected by the DM to build the response (\S\ref{ssec:nlg}),
which is returned to the user and stored as context in the DM.
During the conversation, the DM also communicates with a knowledge graph 
that is stored in the back-end and updated daily by the content management module (\S\ref{ssec:content}).

\subsection{Natural Language Understanding}\label{ssec:nlu}

Given a user's utterance, the NLU module extracts 
the speaker's intent or goals, the desired topic or potential subtopics of
conversation, and the stance or sentiment of a user's reaction to a system
comment.
We store this information in a multidimensional frame which
defines the NLU output. 


To populate the attributes of the frame, the NLU module uses ASR 
hypotheses and the voice user interface output
\cite{Kumar2017NIPS},
as well as the dialog state. 
The dialog state is useful for cases where the system has asked a question with
constraints on the expected response.
A second stage of processing uses parsing results and
dialog state in a set of text classifiers to refine the attributes.

\subsection{Dialog Management}\label{ssec:dm}

We designed the DM according to three high-level objectives: 
 engagement, coherence, and user experience.
The DM takes into account user {\bf engagement} based on components of the NLU output 
and tries to maintain user interest by promoting diversity of interaction strategies (conversation modes).
Each conversation mode is managed by a miniskill
that handles a specific type of conversation segment.
The DM tries to maintain dialog {\bf coherence} by choosing content on the same or a related topic within a conversation
segment,
and it does not present topics or content that were already shared with the user.
To enhance the {\bf user experience}, the DM uses conversation grounding acts to
explain (either explicitly or implicitly) the system's action and to instruct the
user with available options.


The DM uses a hierarchically-structured, state-based dialog model
operating at two levels:
a master that manages the overall conversation, 
and a collection of miniskills
that handle different types of conversation segments. 
This hierarchy enables variety within specific topic segments.
In the Fig.\,\ref{fig:sample_convo} dialog,
Turn\,3 was produced using the \textit{Thoughts} miniskill, 
Turn\,4 using the \textit{Facts} miniskill,
and Turns\,5--7 using the \textit{Movies} miniskill. 
The hierarchical architecture simplifies updating and adding new
capabilities. 
It is also useful for handling high-level conversation mode changes
that are frequent in user interactions with socialbots.

At each conversation turn, a sequence of processing steps are executed to identify a response
strategy that addresses the user's intent and meets the constraints on the
conversation topic, if any.
First, a state-independent processing step checks if the speaker is initiating a new conversation segment (e.g., requesting a new topic). 
If not, a second processing stage executes state-dependent dialog policies.
Both of these processing stages poll miniskills to identify which ones are able
to satisfy constraints of user intent and/or topic. 
Ultimately, the DM produces a list of speech acts and corresponding content 
to be used for NLG, and then updates the dialog state.

\subsection{Natural Language Generation}\label{ssec:nlg}

The NLG module takes as input the speech acts and content
provided by the DM and constructs a response by generating and
combining the response components.

\noindent
{\bf Phrase Generation}:
The response consists of speech acts from four broad categories:
grounding, inform, request, and instruction.
For instance, the system response at Turn\,7 contains three speech acts: grounding
({\it ``Interesting.''}), inform (the IMDB rating), 
and request ({\it ``do you like the movie's director?''}). 
As required by the hosting platform,
the response is split into a message and a reprompt.
The device always reads the message;
the reprompt is optionally used if the device does not detect a response from the user.
The instruction speech acts are usually placed in the reprompt.

\noindent
{\bf Prosody}:
We make extensive use of speech synthesis markup language (SSML) for
prosody and pronunciation to convey information more clearly.
%
to communicate.
We use it to improve the naturalness of concatenated speech acts, to emphasize suggested topics, to deliver jokes more effectively,
to apologize or backchannel in a more natural-sounding way,
and to more appropriately pronounce unusual words.

\noindent
{\bf Utterance Purification}:
The constructed response (which may repeat a user statement) goes through an utterance purifier
that replaces profanity with a non-offensive word chosen randomly from a list of innocuous nouns, often to a humorous effect.

\subsection{Content Management}\label{ssec:content}

Content is stored in a knowledge graph at the back-end, which is updated daily.
The knowledge graph is organized based on miniskills so that 
query
and recommendation can be carried out efficiently by the DM.
The DM drives the conversation forward and generates 
responses by
either traversing links between content nodes associated with the same topic or through relation edges
to content nodes on a relevant new topic.
The relation edges are compiled based on existing knowledge bases (e.g., Wikipedia and IMDB) and entity co-occurrence between content nodes.



Because Sounding Board is accessible to a wide range of users, the system needs to provide content and topics that are appropriate for a general audience.
This requires filtering out inappropriate and controversial material.
Much of this content is removed using regular expressions to catch profanity. 
However, we also  filtered content containing phrases related to sensitive topics
or phrases that were not inherently inappropriate but were often found in
potentially offensive statements (e.g., ``your mother''). 
Content that is not well suited in style to casual conversation (e.g., URLs and lengthy content) is either removed or simplified.

\section{Evaluation and Analysis}
\label{sec:insights}
To analyze system performance, we study conversation data collected from Sounding Board over a one month period
(Nov.~24--Dec.~24, 2017). 
In this period, Sounding Board had 160,210 conversations
with users that lasted 3 or more turns. 
(We omit the shorter sessions, since many involve cases where the user did not intend to invoke the system.)
At the end of each conversation, the Alexa Prize platform collects a rating from
the user by asking {\it ``on a scale of 1 to 5, how do you feel about speaking
with this socialbot again?''} \cite{AlexaPrize2017}.
In this data, 43\% were rated by the user,
with a mean score of 3.65 ($\sigma=1.40$).
Of the rated conversations, 23\% received a score of 1 or 2,
37\% received a score of 3 or 4,
and 40\% received a score of 5.\footnote{Some users give a fractional number
score. These scores are rounded down to the next smallest integer.}
The data are used to analyze how different personality types interact with
the system (\S\ref{ssec:insights_personality})
and length, depth, and breadth characteristics of the conversations
(\S\ref{ssec:insights_content}).


\subsection{Personality Analysis} \label{ssec:insights_personality}


The {\it Personality} miniskill in Sounding Board calibrates user personality
based on the Five Factor model \cite{mccrae1992introduction}
through exchanging answers on a set of personality probing questions 
adapted from the mini-IPIP questionnaire \cite{donnellan2006mini}.


\begin{table}[t]
	\centering
{\small
\begin{tabular}{@{}ll@{\hspace{6pt}}l@{\hspace{6pt}}l@{\hspace{6pt}}l@{\hspace{6pt}}l@{}}
\toprule
               & ope     & con               & ext     & agr     & neu \\ \midrule 
\% users       & 80.02\% & 51.70\%           & 61.59\% & 79.50\% & 42.50\%           \\ \midrule
\# turns       & 0.048** & \textit{not sig.} & 0.075** & 0.085** & \textit{not sig.} \\
rating         & 0.108** & \textit{not sig.} & 0.199** & 0.198** & \textit{not sig.} \\
\bottomrule
\end{tabular}
}%
\caption{
    Association statistics between personality traits
    (\underline{\bf ope}nness, \underline{\bf con}scientiousness, \underline{\bf ext}raversion, 
    \underline{\bf agr}eeableness, \underline{\bf neu}roticism) and $z$-scored conversation metrics.
    ``\% users'' shows the proportion of users scoring positively on a trait.
    ``\# turns'' shows correlation between the trait and the number of turns, and
    ``rating'' the correlation between the trait and the conversation rating, controlled for number of turns.
    Significance level (Holm corrected for multiple comparisons):~$^{**}p<0.001$.
	}
\label{tab:pers_insight}
\end{table}


We present an analysis of how different personality traits interact with Sounding Board, as seen in Table~\ref{tab:pers_insight}.
We find that personality only very slightly correlates with length of conversation (\# turns).
However, when accounting for the number of turns, personality correlates moderately with the conversation rating.
Specifically, we find users who are more extraverted, agreeable, or open to experience tend to rate our socialbot higher.
This falls in line with psychology findings \cite{mccrae1992introduction}, which associate extraversion with talkativeness, agreeableness with cooperativeness, and openness with intellectual curiosity.\footnote{These insights should be taken with a grain of salt, both because the mini-IPIP personality scale has imperfect reliability \cite{donnellan2006mini} and user responses in such a casual scenario can be noisy.}



\subsection{Content Analysis} \label{ssec:insights_content}

\begin{figure}
	\centering 
	\includegraphics[width=0.48\textwidth,trim=1cm 1cm 1cm 1cm]{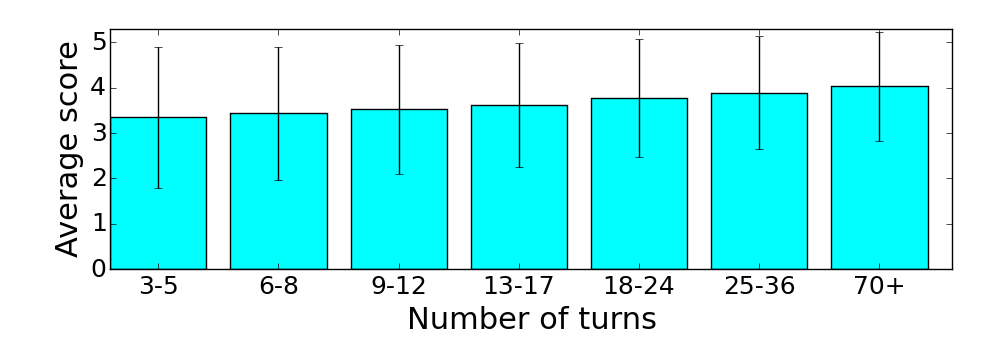}
	\caption{Average conversation score by conversation length. 
	Each bar represents conversations that contain the number of turns in the
	range listed beneath them and is marked with the standard deviation.}
	\label{fig:avg_score}
\end{figure}

Most Sounding Board conversations were short (43\% consist of fewer than 10
turns), but the length distribution has a long tail. The longest
conversation consisted of 772 turns, and the average conversation length was
19.4 turns.
As seen in Fig. \ref{fig:avg_score}, longer conversations tended to
get higher ratings.

While conversation length is an important factor,
it alone is not enough to assess the conversation quality, as evidenced by the low correlation with user ratings ($r=0.14$) 
and because some turns (e.g., repairs) 
may have a negative impact.
Therefore, we also study the breadth and depth of the sub-dialogs 
within conversations of roughly equal length (36--50) with high (5) vs.~low (1--2) ratings.
We automatically segment the conversations into sub-dialogs based on the system-identified topic, and annotate each sub-dialog as engaged or not depending on the number of turns where the system detects that the user is engaged. The breadth of the conversation can be roughly characterized by the number and percentage of engaged sub-dialogs; depth is characterized by the average number of turns in a sub-dialog.
We found that the average topic engagement percentages differ significantly (62.5\% for high scoring vs.~28.6\% for low), but the number of engaged sub-dialogs were similar (4.2 for high vs.~4.1 for low). Consistent with this, the average depth of the sub-dialog was higher for the high conversations (4.0 vs.~3.8 turns).





\section{Conclusion}
\label{sec:conclusion}
We presented Sounding Board, a social chatbot that has won the inaugural Alexa Prize Challenge. As key design principles, our system focuses on providing conversation experience that is both user-centric and content-driven. 
%
%
Potential avenues for future research include increasing the success rate of the topic suggestion and improving the engagements via better analysis of user personality  and topic-engagement patterns across users.





\section*{Acknowledgements}
In addition to the Alexa Prize financial and cloud computing support, 
this work was supported in part by NSF Graduate Research Fellowship (awarded to E. Clark), NSF (IIS-1524371), and DARPA CwC program through ARO (W911NF-15-1-0543). 
The conclusions and findings are those of the authors and do not necessarily reflect the views of sponsors.

\newpage
\bibliographystyle{acl_natbib}
\bibliography{alexaprize_ref}

\end{document}